\documentclass[letterpaper, 12pt]{article}
\usepackage{latexsym}
\usepackage{epsfig,amssymb,euscript}
\usepackage{amsmath}
\usepackage{mathrsfs}
\usepackage{color}

\usepackage{pstricks}

\usepackage[margin=30pt,font=small,labelfont=bf]{caption}
\numberwithin{equation}{section}  
\newsavebox{\ns}
\newsavebox{\dbrane}

\def\be{\begin{equation}}
\def\ee{\end{equation}}
\def\bea{\begin{eqnarray}}
\def\eea{\end{eqnarray}}

\renewcommand{\theequation}{\arabic{section}.\arabic{equation}}
\def\theequation{\thesection.\arabic{equation}}

\newcommand{\cQ}{{\cal Q}}

\def\Dslash{\,\,{\raise.15ex\hbox{/}\mkern-12mu D}}
\def\Dbarslash{\,\,{\raise.15ex\hbox{/}\mkern-12mu {\bar D}}}
\def\delslash{\,\,{\raise.15ex\hbox{/}\mkern-9mu \partial}}
\def\delbarslash{\,\,{\raise.15ex\hbox{/}\mkern-9mu {\bar\partial}}}
\def\pslash{\,\,{\raise.15ex\hbox{/}\mkern-9mu p}}
\def\calDslash{\,\,{\raise.15ex\hbox{/}\mkern-12mu {\cal D}}}

\def\spaQ{\mathbf{Q}}

\newcommand{\nn}{\nonumber \\}
\newcommand{\reef}[1]{(\ref{#1})}

\begin{document}

\makeatletter
\renewcommand{\theequation}{\thesection.\arabic{equation}}
\@addtoreset{equation}{section}
\makeatother

\baselineskip 18pt

\begin{titlepage}

\vfill

\begin{flushright}
MIT/CTP 4455\\
\end{flushright}

\vfill

\begin{center}
   \baselineskip=16pt
   {\Large\bf  On universality of charge transport in AdS/CFT}
  \vskip 1.5cm
       Julian Sonner\\
   \vskip .6cm
   \begin{small}
      \textit{Center for Theoretical Physics, Massachusetts Institute of Technology\\ Cambridge, MA 02139, U.S.A.}
        \end{small}\\*[.6cm]

   \begin{small}
      \textit{and}
        \end{small}\\*[.6cm]
           \begin{small}
      \textit{L.N.S., Massachusetts Institute of Technology\\ Cambridge, MA 02139, U.S.A.}
        \end{small}\\*[.6cm]
        \vskip .6cm
\end{center}

\vfill

\begin{center}
\textbf{Abstract}
\end{center}

\begin{quote}
We develop the holographic formulation of transport in strongly coupled two-layer systems. We identify a  dc conductivity, $\sigma^{\rm dc}$, that is finite even in a translationally invariant setup, and universal for CFTs with a gravity dual. The thermoelectric conductivity and heat conductivity are fully determined by the electrical conductivity matrix, as a consequence of Ward identities. We use the memory-matrix approach for double-layer systems, together with Ward identities, to show that $\sigma^{\rm dc}$ - extended to finite frequency - has no Drude peak and, similarly, that its universal value is unaffected if translation invariance is softly broken.
\end{quote}

\vskip1em
\begin{center}
\end{center}

\vfill

\end{titlepage}
 

\section{Introduction}
Charge transport is a fundamental manifestation of the underlying physical principles of any condensed-matter system and thus a quantity of intense interest in any given theory or experiment. In a field-theory approach, transport is characterized by near-equilibrium correlation functions with lorentzian signature. At strong coupling these correlation functions are often inaccessible, due to the reliance on perturbative methods and due to severe restrictions of lattice simulations for finite-density quantities that are intrinsically defined in lorentzian signature. It is generally accepted as an attractive feature of the AdS/CFT correspondence \cite{Maldacena:1997re,Witten:1998qj,Gubser:1998bc} that it provides a framework for the computation of transport quantities for strongly-interacting field theories \cite{Son:2002sd}. This approach is most promising for transport coefficients that are universal within models with a gravity dual. The most well-known such transport coefficient is the ratio of shear viscosity over entropy density, $\eta/s$ of a relativistic strongly coupled plasma with a gravity dual, which takes on the universal value of $1/4\pi$ in suitable units \cite{Policastro:2001yc}. Such quantities, however, are hard to come by. It is of great interest and importance to continue the search for transport coefficients and universal mechanisms if we are to push the applicability of holographic methods further towards realistic systems. Applying the techniques of holography to condensed matter phenomena is a promising arena in which to look for such universal transport effects. There is, however an immediate obstruction when we consider the properties of theories at finite charge density. The most directly accessible quantities, namely dc charge-transport coefficients, are ill defined in the continuum limit, since they are strictly divergent. The divergence is directly linked to translation invariance; its breaking - either by impurities or via lattice effects - is likely not universal. It is nevertheless sometimes possible, for example by exploiting the hydrodynamic nature \cite{PhysRevLett.85.1092,Roscharxiv,PhysRevB.68.104401} of the late-time transport, to extract scaling laws \cite{Hartnoll:2007ih,Hartnoll:2012rj,Donos:2012ra,Mahajan:2013cja} for dc conductivities. These mechanisms rely on a finite Drude weight in the conductivity, associated with approximately conserved quantities. In this paper we take a different angle, focusing on a quantity with vanishing Drude weight and derive universal results which do {\it not} involve hydrodynamic modes and can therefore be considered intrinsic to the microscopic field theory.

To this end we introduce a new transport coefficient, {\it transconductance} - or its inverse {\it transresistance} - into the holographic dictionary and show that a suitably defined transport coefficient characterizing transconductance is finite even in the translationally invariant setup, and in fact turns out to take a universal value for (2+1) dimensional CFTs. The dc transconductance exhibits complete bulk universality and has the value
\be\label{eq.universal}
\sigma^{\rm dc} = \frac{1}{e^2_{\rm eff}}\,,
\ee
in any $2+1$ dimensional quantum field theory with a gravity dual that asymptotes to $AdS_4$. The effective bulk coupling $e^2_{\rm eff}$ is related to the number of charged degrees of freedom, i.e. a central charge, of the IR fixed point of this theory. There is a well-known exception to our earlier statement that dc charge transport must be infinite in a translationally invariant setup  \cite{PhysRevB.56.8714}: a particle-hole symmetric system (such as a zero-density CFT) can give rise to a finite current, because a finite charge current involving particles and holes in equal numbers carries no net momentum. The transport we are describing here is a realization of an analogous mechanism, but applies away from the particle-hole symmetric point. In other words the transport associated with the {\it universal} coefficient \reef{eq.universal} involves a charge flow with no associated net momentum flow and is thus finite for a clean system. As far as we know this is the first time such a mechanism is pointed out in a system that has a net charge density and is not at the particle-hole symmetric point. It likely has implications for other systems and approaches, holographic or not.

One can furthermore use the memory matrix approach to show that the universal value of $\sigma^{\rm dc}$ only receives parametrically small corrections if translation invariance is broken via a coupling to the leading irrelevant symmetry-breaking operator, ensuring that the theory still flows to the same IR fixed point.  Although our approach applies to quantum critical dynamics, we expect our results to be more widely valid, indeed they may well follow from an analysis of the constraints placed upon our system by relativistic hydrodynamics, for example within a Boltzmann approach \cite{PhysRevB.56.8714}.  It has been  suggested that clean graphene monolayers can be thought of as a relativistic critical system \cite{MuellerSachdevFritz,PhysRevB.78.085416}, at least for a window of temperatures where the effective speed of light runs very little with scale, so our approach to the physics of double layers may be of direct relevance to the case of graphene. For a probe-brane holographic approach to double-layer graphene see \cite{Grignani:2012qz}.

This paper is structured as follows. In the remainder of this introduction we give a definition of transconductance, the transport at the focus of this work. Section \ref{sec.universal_transport} introduces the general class of holographic bottom-up models which exhibit transconductance and uses general arguments to establish Eq.  \reef{eq.universal}. Section \ref{sec.condmatrix} serves to anchor $\sigma^{\rm dc}$ in a more general discussion of the conductivity matrix of the models introduced in Section \ref{sec.universal_transport}  and then discusses constraints placed on transport by Ward identities. Section \ref{sec.momentum_relax} considers the effect of translation breaking via impurities or lattice effects and uses the memory function matrix approach (with a short review) together with the clean Ward identities to establish general forms of low-frequency limits of transconductance. The conclusions are followed by a short appendix on some details of the holographic renormalization of our action.
\subsection{Introducing: a new transport coefficient\label{sec.transconductance}}

Transresistance, conceptually, measures the amount of drag between charge carriers in two different layers\footnote{We are interested in this paper in two-dimensional sheets or layers, each governed by a 2+1 strongly-interacting field theory.}, under the condition that there is no direct interaction between them. Systems where such a quantity can be investigated typically have two conducting layers separated by an insulating layer, for example two sheets of graphene sandwiching a sheet of boron nitride \cite{GeimGroupExperiment}. One often refers to the whole setup as a {\it heterostructure}. The case where the spacing between the layers is so small that quantum-mechanical wave-function spread means that the whole heterostructure can be viewed as a two-dimensional system is particularly interesting and is the focus of this work.

\begin{figure}[h!]
\begin{center}
\includegraphics[width=0.4\textwidth]{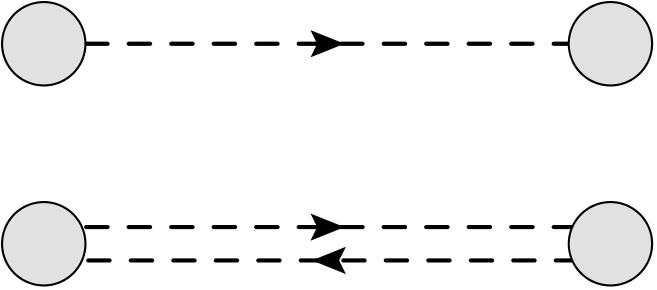}
\begin{picture}(0.1,0.25)(0,0)
\put(-9.6,4.2){\makebox(-330,14){$J^{(-)}$}}
\put(-9.6,4.2){\makebox(-330,128){$J^{(1)}$}}
\put(-9.6,4.2){\makebox(-10,14){$J^{(-)}$}}
\put(-9.6,4.2){\makebox(-10,128){$J^{(2)}$}}
\put(-9.6,4.2){\makebox(-160,45){$P$}}
\put(-9.6,4.2){\makebox(-160,145){$P$}}
\end{picture}\hskip3em\includegraphics[width=0.4\textwidth]{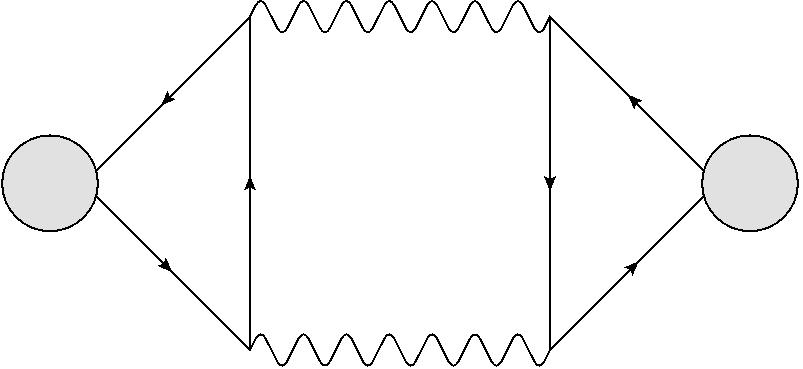}\begin{picture}(0.1,0.25)(0,0)
\put(-9.6,4.2){\makebox(-324,80){$J^{(1)}$}}
\put(-9.6,4.2){\makebox(-5,80){$J^{(2)}$}}
\end{picture}
\caption{a) The leading contribution to drag in a holographic computation. The two currents interact via mixing with the momentum operator $T_{0x}:=P$. The diagram contributing to the finite quantity $\sigma^{\rm dc}$ involves no net momentum transfer, as indicated by the two opposite momentum-flow arrows. b) Example of contribution in perturbative formulation of drag. The currents have no direct interaction, but talk to each other via loops in each layer that in turn couple through e.g. the Coulomb interaction (wiggly lines). The diagram with only a single wiggly line joining the two fermion loops vanishes at $\mathbf{Q}=0$.
\label{fig:diagrams}}
\end{center}
\end{figure}
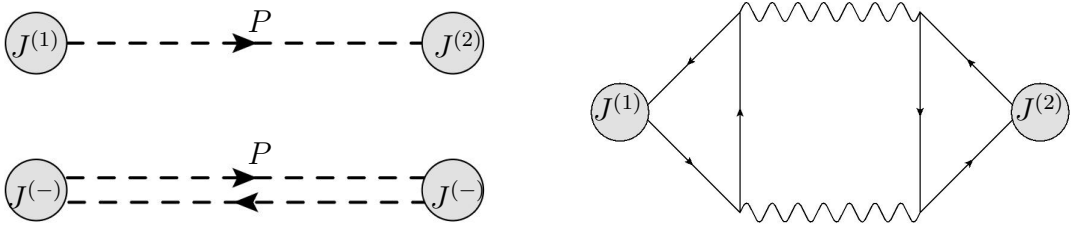

Drag is defined as follows: One measures the induced voltage in one layer due to an applied current in the other. More precisely, one applies a current $J_1^a$ in layer one (the `active' layer), and then measures the induced voltage $V_2^b$ in layer two, the `passive' layer, under the condition that there is no current flowing in the passive layer. This phenomenon is often referred to as  {\it Coulomb drag}, since in many systems, such as graphene heterostructures or Hall double layers, interlayer interactions are mediated by the Coulomb force (see Fig. \ref{fig:diagrams}).

 In a field-theory treatment of this phenomenon the basic object which one needs to determine is the conductivity matrix $\sigma_{ab}^{ij}$, defined in terms of a Kubo formula  \cite{Mahan,KamenevAndOreg} as
\be\label{eq:Kubo}
\sigma_{ab}^{ij}(\omega,\mathbf{Q}) = \frac{1}{{\cal A}\,\omega} \int dt\, \theta(t)e^{i\omega t} \langle \left[ J_a^i(t,\mathbf{Q})\,,\, J_b^j (0,\mathbf{Q}) \right] \rangle\,.
\ee
In this expression ${\cal A}$ is the area of the sample, $i,j$ are spatial indices and $a,b$ label the individual layers. The integrand is nothing other than the retarded correlation function of the conserved currents in both layers 
\be
G_{ab}^{ij,\,R}(t\,,\spaQ) =  -i \theta(t)\langle \left[ J_a^i(t,\spaQ)\,,\, J_b^j (0,\spaQ) \right] \rangle\,.
\ee
In this paper we use holography to calculate this correlation function and thus the transconductance in a strongly-coupled double-layer system. The conductivity associated with a single $U(1)$ sector was the object of much attention (see e.g. \cite{Herzog:2007ij,Hartnoll:2007ai,Hartnoll:2007ip,Kovtun:2008kx,Hartnoll:2008hs})  and we shall use these results as a benchmark for the ones here. In this paper we are interested in the conductivities themselves and so we will mostly study the correlation functions at zero momentum $\spaQ=0$. Of course the full correlation function \reef{eq:Kubo} contains more physical information, which will be useful, for example, in developing a hydrodynamic description of drag.
In perturbative treatments of drag, one often has the restriction $\sigma_{12}\ll \sigma_{11}, \sigma_{22}$ (see e.g. \cite{KamenevAndOreg}), i.e. one assumes that the interlayer drag is very weak. In this work we shall have no such restriction, and we are able to analyze the case where the drag itself is as strong as the intralayer interactions. These interactions are those of a strongly-coupled field theory.

\section{Finite dc transport with translation invariance\label{sec.universal_transport}}
From the discussion above we can see that the requirements on our model are simple. Each of the conserved currents in the theory gives rise to a local $U(1)$ gauge field $A_{(i)}$ in the bulk. Since we do not want to consider situations where the two currents have a direct interaction, we do not include a cross kinetic term for the two gauge fields in the bulk, which in principle would be allowed by the $U(1)\times U(1)$ local symmetry. Generic supergravity models usually have gauge kinetic terms parametrized by a (neutral) scalar field - we refer to this as a `dilaton' even though it is not {\it the} stringy dilaton field. Hence we also include a field $\varphi$ with a potential $V(\varphi)$. We shall not need the exact form of the potential or the function $\Phi[ \varphi]$ that governs the coupling of the dilaton to the gauge fields, for our  argument, but we should require that the field $\varphi$ has at least marginal fall off as one approaches the boundary and that, $\Phi(z) \rightarrow 1$ in this limit. Indeed in this paper, we are concerned with generic quantities and mechanisms, which will be necessarily satisfied in any specific microscopic realization.
Since we consider both of our layers as equivalent physical systems individually, their conserved currents necessarily couple equally to the rest of the theory. As a consequence the effective bulk description is of the form
\be\label{eq:bilayer_actionI}
S = \int\sqrt{|g|}\left(\frac{1}{2\kappa^2}\left( R + \frac{6}{\ell^2} -  (\partial\varphi)^2 - 2 V(\varphi)\right) - \frac{\Phi[\varphi]}{4e^2}\left(F_{(1)}^2 +F_{(2)}^2\right) \right)d^4 x
\ee
and we see that it has an $SO(2)$ symmetry rotating the gauge fields into one another. We can think of this as a flavor symmetry of currents. If  only a layer-exchange $\mathbb{Z}_2$ symmetry remains, one could allow more general couplings than those in \reef{eq:bilayer_actionI}, in particular each layer could then couple differently to the remaining degrees of freedom. Nevertheless, the remaining exchange symmetry is still sufficient to argue finiteness of $\sigma^{\rm dc}$, as we shall see in section \ref{sec.secdrag}. Holographic theories with two bulk Maxwell fields have been considered previously both with `conventional' AdS asymptotics \cite{Bigazzi:2011ak,Musso:2013ija}, and with a few on non-relativistic holography \cite{Smolic:2013gx}.

 In section \ref{sec:Ward} we will add a suitable counterterm action, which makes the variational principle well defined and holographically renormalizes the dual theory. The action \reef{eq:bilayer_actionI} does not contain any couplings between $F_{(1)}$ and $F_{(2)}$, so one might think that there can be no interlayer transport. However, as we shall see, interlayer transport is mediated by a mixing of the currents with the momentum operator $P = T_{tx}$, which is holographically dual to a fluctuation of the metric, $\delta g_{tx}$. Both gauge fields are coupled to the metric, and so the leading-order contribution to drag in our setups proceeds via a momentum exchange (graviton exchange) as illustrated in Fig \ref{fig:diagrams}a. Before turning to the computation of the conductivities, however, we need to establish some facts about the background solutions appropriate for the physics we are interested in.
\subsection{Two-charge solutions and their flavor rotation}
The equilibrium ensemble of the double-layer system is described by a two-charge (dilaton) black hole, a solution of the theory \reef{eq:bilayer_actionI} of the form
\bea\label{eq:background_ansatz}
\varphi &=& \varphi(z)\,,\nn
A_{(i)} &=& \phi_{(i)}(z)dt\,,\qquad i = 1,2\nn
ds^2 &=& G_{\mu\nu}dx^\mu dx^\nu=\frac{\ell^2}{z^2} \left(-f(z)dt^2 + f(z)^{-1}dz^2 + d{\mathbf{x}^2}\right)\,.
\eea
Explicit examples of $AdS$ dilaton black holes \cite{Goldstein:2009cv,Goldstein:2010aw}, and their IR limits \cite{Charmousis:2010zz} have recently been considered in the holographic literature  but for the purpose of the analysis here it is sufficient to suppose that a solution can be constructed (numerally or analytically). Our arguments are made on a formal level and do not rely on any particular solution.
Consider now the gauge-field equations
\be
\nabla_\mu \left( \Phi[\varphi] F^{\mu}_{(i)}{}_\nu \right)=0\,.
\ee
Plugging in the ansatz \reef{eq:background_ansatz} one deduces that these have a first integral
\be
\left( \Phi[\varphi] \phi_{(i)}' \right)' = 0\,\qquad \phi_{(i)}' (z) = \frac{  C_{(i)}}{\Phi(z)}\,.
\ee
This can be integrated to give the formal solution
\be\label{eq:formgauge}
\phi_{(i)} = \int^z_{z_h} \frac{C_{(i)}d\tilde z}{\Phi(\tilde z)}\,,
\ee
where the lower limit of integration is chosen so that $A_t(z_h)=0$. It is useful to define the dimensionless integral
\be
I_\Phi=\int_0^{1} \frac{d\zeta}{\Phi(\zeta)}\,,\quad (\zeta = z/z_h)\,,
\ee
which integrates the flow of the coupling constant as a function of the bulk coordinate $z$. This integral may have an infra-red divergence depending on the details of the solution. If the dilaton coupling function is trivial, we have simply $I_\Phi=1$.
The constants $C_{(i)}$ are the charge densities\footnote{The sign is convention, coming from the fact that in AdS/CFT the gauge field is dual to a boundary current $\langle J^\alpha_{(i)} \rangle \sim F^{\alpha}_{(i)}{}_z$ so that the charge density $\langle J^0_{(i)} \rangle$ comes with a timelike index up accounting for the negative sign in \reef{eq.hodgecharge}.}  associated with the gauge fields, as can be seen from the integral
\be\label{eq.hodgecharge}
\langle n_i \rangle = -\frac{1}{2e^2{\cal A}} \int \star\, \Phi[\varphi] F_{(i)} = - \frac{C_{(i)}}{2e^2}\,,
\ee
where $\star$ denotes the Hodge dual and ${\cal A}$ is the area of the spatial $\mathbb{R}^2$ of the dual field theory as in \reef{eq:Kubo}. This is equivalent to defining the charges by differentiating the action with respect to the boundary value of $\phi_{(i)}$.  The boundary limit of the expression \reef{eq:formgauge} in fact defines the chemical potentials, which can be brought into the form
\be
\mu_i = 2 e^2 n_i z_h I_\Phi\,.
\ee
This  allows us to determine the charge susceptibility $\chi_{ij} = \frac{\partial n_i}{\partial \mu_j}\Bigr|_{n_i=0} = \chi(T)\delta_{ij}$. One computes
\be
\chi(T)= \frac{1}{2e^2 z_h(T) I_\Phi}\,.
\ee
The dependence on temperature is left implicit and is hidden in the horizon radius $z_h(T)$, evaluated at vanishing charge density. If we now define the quantity $\bar\mu = C_{(2)}/C_{(1)}$ we can construct the $SO(2)$ matrix
\be
{\cal M}(\bar\mu) = \frac{1}{\sqrt{1 + \bar\mu^2}} \left(\begin{array}{cc}1 & \bar\mu \\ -\bar\mu & 1\end{array}\right)\,,
\ee
which rotates the gauge fields into a new basis
\be
\left(\begin{array}{c}A_+ \\A_-\end{array}\right) = {\cal M}(\bar\mu) \left(\begin{array}{c}A_{(1)} \\ A_{(2)}\end{array}\right)\,,
\ee
and acts on the charges as $C_+ = C_{(1)} (1 + \bar\mu^2)^{1/2}$ and $C_-=0$. The action on the field strengths similarly gives
\be
F_{-} =0\,,\qquad F_{+} = \frac{C_{+}}{\Phi(z)}dz\wedge dt\,.
\ee
Hence the black hole is charged under only one combination of the fields in the new basis. In order to determine charge transport of this system, we need to solve the quadratic fluctuation equations, expanded around the background \reef{eq:background_ansatz}. A little thought then shows that the fluctuations of $A_-$ cannot couple to any other fluctuations of the system, i.e. to fluctuations of the metric, the other gauge field, or the dilaton, for the simple reason that these couplings, for example on the level of the action, have to be of the form $F_- \delta A_- \delta \alpha$. We have denoted a generic fluctuation of the other fields as $\delta \alpha$. Since the background has $F_{-} =0$, no such couplings can arise. We emphasize that this result is completely general, i.e. it holds for a general perturbation $\delta \alpha (t,z,\mathbf{x})$ with arbitrary time and space dependence.
\subsection{Universal charge transport\label{sec:finite_transport} and its finite dc limit}
In order to determine the dc transport of the model we need to determine the two-point correlation functions of the current dual to the gauge fields $A_{(i)}$. For this one needs the quadratic fluctuation action.
We can summarize the discussion above by stating that we are now faced with solving the charge transport properties of the current $J_-$, dual to the bulk field $A_-$ with an effective theory of the form
\be\label{eq.fluctuationaction}
S_{\rm eff} = - \frac{1}{4e^2} \int\sqrt{g}\Phi[\varphi] F_{-}^2 + S[g,F_+,\varphi]\,,
\ee
which experiences a background that is {\it neutral} with respect to the field $A_-$. This implies that quadratic fluctuations of $A_-$ completely decouple from the ones of $S[g,F_+,\varphi]$ and we only need to consider the first term in \reef{eq.fluctuationaction}.  It then follows directly that the membrane-paradigm results of \cite{Iqbal:2008by} can be applied to the theory at hand. In the present context it means that there is a {\it universal} dc conductivity associated with $A_-$, namely
\be\label{eq:finitedc}
\sigma^{\rm dc} = \frac{1}{e^2 (z_{\rm IR})}\,,
\ee
where one defines the scale-dependent gauge coupling $e^{-2}(z) = \Phi(z)/e^2$ and $z_{\rm IR}$ is the location of the infrared horizon. Alternatively one can derive this result directly from the fluctuation equations following from \reef{eq.fluctuationaction}. What matters is that the answer is universal from the perspective of the bulk dual. Unfortunately, it is not model-independent from the point of view of the dual field theory due to the dependence on the dilaton profile through the scale-dependent coupling $e(z)$. However, for a large class of cases it is in fact universal from the field-theory perspective, namely for any CFT$_3$, where it is determined by the zero-temperature current-current correlator \cite{Freedman:1998tz,Kovtun:2008kx}
\be\label{eq.currentcurrent}
\langle J_{\mu}(x) , J_\nu(0) \rangle = \frac{k}{x^4} \left( \delta_{\mu\nu}  - 2 \frac{x_\mu x_\nu}{x^2} \right)\,,
\ee
with $k = \frac{2}{\pi^2 e^2}$, so that $\sigma^{\rm dc}({\rm CFT}_3) = \frac{\pi^2k}{2}$. The quantity $k$ is  a dimensionless central charge of the CFT for charged degrees of freedom. This is a general result, since we did not  have to use any of the details of the solution of \reef{eq:background_ansatz}, which will strongly depend on whatever theory we want to solve. Moreover, such a solution, typically will not be analytically tractable. Even if the dilaton has a nontrivial profile, the effective coupling constant $e^2(z_{\rm IR})$ can still be related to a central charge at the IR fixed point, but in this case the central charge may depend on the chemical potentials and the correlator \reef{eq.currentcurrent} will have to be evaluated in the IR theory. Dividing the conductivity by the charge susceptibility\footnote{This ratio has previously been obtained by \cite{Kovtun:2008kx} for a theory without dilatonic couplings and our result agrees with theirs in this case.},  we obtain the simple expression
 \be\label{eq.chargeoversusceptibility}
 \frac{\sigma^{\rm dc}}{\chi} =2 z_h(T) \Phi(z_h) I_\Phi\,.
 \ee
We have thus identified a transport coefficient, $\sigma^{\rm dc}$ which is {\it finite} in the dc limit, without having to introduce a mechanism of momentum relaxation. Furthermore, we showed that the decoupling of the dynamics of the field dual to the transport current giving rise to $\sigma^{\rm dc}$ is completely general, and in particular extends to finite-momentum fluctuations. This also implies that the value of $\sigma^{\rm dc}$ is unaffected by the detailed mechanism of momentum relaxation, such as weak impurity scattering or lattice effects, simply because the current operator $J_-$ has zero overlap with the  momentum operator. We shall return to this in later sections.
  
Finally we should note that the derivation above focused on the dc conductivity, but in the case of certain CFTs with a bulk EM duality \cite{Herzog:2007ij}, the result \reef{eq:finitedc} in fact extends to the optical conductivity, meaning that it is independent of frequency and takes the form
 \be\label{eq.optical}
 \sigma^{\rm drag}(\omega) = \frac{\pi^2 k}{2}\,.
 \ee
 Relations of this kind have been noted earlier for  the optical conductivity of a single current, for example  by \cite{Kovtun:2008kx,Herzog:2007ij}. In such cases, however, one actually gets a divergent dc limit due to translation invariance unless the background is neutral. Then, since the current has a non-vanishing overlap with the momentum operator, the mechanism to relax momentum will not only resolve this divergence into a Drude peak, but also destroy the frequency independence of the optical conductivity. This is in contrast with our case, where the relevant current operator has zero overlap with the momentum operator even at finite density. Including higher-derivative terms in the bulk action, breaking the self-duality of the Maxwell term, is also be expected to modify the frequency-independence of the optical conductivity, as was shown in the case of a single current in \cite{Myers:2010pk}.
 \section{The (rest of the) conductivity matrix\label{sec.condmatrix}}
We saw in the section above that the difference of the fluctuations $A_-$ sees an uncharged black hole, while the average of the fluctuations $A_+$ couples to the sum of the charges. As a consequence of this decoupling the retarded current-current\footnote{Note that there are other non-vanishing correlators, namely those between stress-tensor and current $ \langle \left[ T_{0x},J_\pm \right]\rangle = G_{T\pm}^R$ and stress-tensor and stress-tensor $ \langle \left[ T_{0x},T_{0x} \right]\rangle = G_{T T}^R$. We will see that they are entirely determined in terms of the current-current correlations functions.} correlation functions of the dual field theory are diagonal in this basis :
\bea
\langle \left[ J_\pm , J_\mp \right]\rangle &=& 0\nn
\langle \left[ J_\pm , J_\pm \right]\rangle &=&G^R_{\pm\pm}
\eea
We would like to relate this to the conductivity matrix, $\sigma_{ij}$, as defined in the original basis. It is easy to convert the corresponding expressions:
\bea\label{eq:sigmachangeofbasis}
\sigma_{11}&=&- \frac{1}{(1 +  \bar\mu^2)} \frac{1}{\omega}\left(\langle \left[ J_+, J_+ \right] \rangle  + \bar\mu^2 \langle\left[ J_- , J_-\right] \rangle\right)\nn
\sigma_{12}&=&- \frac{\bar\mu}{(1 +  \bar\mu^2)} \frac{1}{\omega}\left(\langle \left[ J_+, J_+ \right] \rangle  - \langle\left[ J_- , J_-\right] \rangle\right) = \sigma_{21}\nn
\sigma_{22}&=&- \frac{1}{(1 +  \bar\mu^2)} \frac{1}{\omega}\left(\bar\mu^2\langle \left[ J_+, J_+ \right] \rangle  + \langle\left[ J_- , J_-\right] \rangle\right)\,.
\eea
One can see that each element of the conductance matrix in this basis gets contributions both from $J_+$ and from $J_-$, and each matrix element thus picks up a dc divergence from the $J_+$ part. Moreover, our arguments above established that the universal transport quantity in this system is associated with the $\langle  \left[ J_-, J_- \right]\rangle$ correlator. This is a physically accessible quantity, namely
\be\label{eq:drag_def}
\sigma^{\rm drag}(\omega) = \sigma_{12}(\omega) - \bar\mu^{-1} \sigma_{11}(\omega)\,.
\ee
so that
\be
\sigma^{\rm dc} = \lim_{\omega\rightarrow 0}\sigma^{\rm drag}(\omega)\,.
\ee
This definition makes physical sense. Suppose we want to learn about the amount of drag between the layers by measuring the `bare' coefficient $\sigma_{12}$. This will get contributions both from the momentum transferred between the layers (the drag we are interested in) and the momentum dissipated due to interlayer momentum relaxation. If no such interlayer momentum relaxation is effective, we instead measure a divergence in the dc transport associated with the interlayer translation invariance. It is thus natural to focus on the subtracted quantity \reef{eq:drag_def}. Of course, the entire point of section \ref{sec:finite_transport} was that there is an independent reason to focus on \reef{eq:drag_def}, namely that it quantifies a universal transport property of strongly coupled field theories with gravity duals.
Typically one is interested in the equal charge case, $\bar\mu=1$, so that the definition $\sigma^{\rm drag} = \sigma_{12}-\sigma_{11}$ is  independent of the densities and $\sigma_{12} = \frac{i}{2\omega} \left( G_{--} - G_{++} \right)$ etc. are particularly simple. 
 \subsection{Global symmetries and gauge fixing}
In linear response the conductivity is encoded in the retarded correlation function of currents at zero momentum via the Kubo formula \reef{eq:Kubo}:
\be\label{eq:interlayer_Kubo}
\sigma_{ab}^{ik}(\omega) = \frac{i G^{ik}_{ab}{}^R(\omega\,, \mathbf{Q}=0)}{{\cal A}\,\omega}\,.
\ee
As indicated  the correlation function is evaluated at vanishing spatial momentum. In situations where parity is broken, as for example in the presence of a magnetic field, there can be an off-diagonal Hall contribution to the conductivity. The remaining continuous symmetries at $\mathbf{Q}=0$ then dictate
\be
\sigma^{ij}_{ab} = \sigma^{ij}\delta_{ab} + \Sigma^{ij}\epsilon_{ab}\,,
\ee
where $\sigma^{ij}$ is the direct conductivity and $\Sigma^{ij}$ is the Hall contribution. Off-diagonal elements in layer indices of the conductivity matrix are referred to as trans- or drag- conductivity.
We have already seen that that $\sigma_{12} = \sigma_{21}$. By the transverse $SO(2)$ symmetry we can calculate $\sigma_{12}$ by considering fluctuations only in the $x$ direction\footnote{In the absence of a magnetic field, or another mechanism of breaking $P$, the Hall transconductance vanishes.}.
At vanishing spatial momentum the spin-0 dilaton fluctuation cannot mix to linear order with the spin-1 fluctuations of the metric which we need to consider. In order to calculate the quantity \reef{eq:interlayer_Kubo} we thus only consider vector fluctuations of the metric and gauge fields, of which there are four. The deviations of the metric and gauge fields from their vanishing background (see \reef{eq:background_ansatz}) are parametrized by
\be\label{eq:fluctuations}
\left\{ A_x^{(i)}(t,z)\,,g_{tx}(t,z)\,, g_{zx}(t,z) \right\}\,.
\ee
More precisely these functions are related to small fluctuations of metric and gauge field as follows:
\bea
\delta G_{tx}(t,z) &=& \frac{\ell^2}{z^2}g_{tx}(t,z)\nn
\delta G_{zx}(t,z) &=& \frac{\ell^2}{z^2}g_{zx}(t,z)\nn
\delta A_{x}^{(i)}(t,z) &=&A_{x}^{(i)}(t,z)\,.
\eea
We will work with their Fourier transforms
\be
 A_x^{(i)}(t,z) = \int\frac{d\omega}{2\pi}e^{-i \omega t}A_x^{(i)}(\omega,z)\,,\qquad  g_{tx}(t,z) = \int\frac{d\omega}{2\pi}e^{-i \omega t}g_{tx}(\omega,z).
\ee
For the problem at hand, it turns out that a convenient choice of gauge sets $g_{zx}(\omega,z)=0$. This `radial' gauge does not fix the diffeomorphism symmetry completely, so that there is a residual gauge transformation
\be\label{eq:residualgauge}
{\cal L}_\xi g_{tx} = -i \omega \xi^x\,,\qquad {\cal L}_\xi A_x^{(i)} = {\cal L}_\xi g_{zx}=0 
\ee
for constant gauge parameter $\xi^x$. This takes the form of an infinitesimal translation in the $x$ direction in the field theory and, as we will see later, is tied to translation invariance of the boundary theory. Note we only need to concern ourselves with the gauge fixing/freedom in the vector sector of which we have just given a complete description. Obviously there is more freedom in the spin-0 and spin-2 sectors which do not mix with the quantities here at zero momentum.

\subsection{Boundary action, correlation functions $\&$ Ward identities}\label{sec:Ward}
Correlation functions in AdS/CFT can be calculated by identifying the on-shell action as the generating functional of correlation functions in the dual theory. In our case, the on-shell action, suitably renormalized, reduces to a boundary term.
Before we can identify the boundary action we must specify the counter term action further. We take the usual form \cite{Balasubramanian:1999re,Skenderis:2002wp}
\be\label{eq.counter}
S_{\rm CT} = \frac{1}{\kappa^2}\int \sqrt{-\gamma}\left({\cal K} + \frac{2}{\ell} \right)d^3x+S_{\rm CT}^\varphi\,.
\ee
The counter term for the dilaton depends on its conformal dimension and thus the quadratic term in $V(\varphi)$, but we will not need its precise form here. For completeness we carry out the holographic renormalisation of the background in full generality in appendix \ref{app.holorenorm}. Remember that we did not specify the details of this fall-off, only that it be at least constant or faster.
Let us now expand the action \reef{eq:bilayer_actionI} including the counter terms to quadratic order in the fluctuations. After integrating by parts the Euclidean renormalized on-shell action $S^{\rm OS} + S_{\rm CT}$ reduces to the finite boundary term
\bea\label{eq:boundaryaction}
W[g_{tx}^0,A^{(i),0}_x]&=&\lim_{z\rightarrow 0} \int\frac{d\omega d^2x}{2\pi}\left[ \sum_i \langle n^{(i)}\rangle   g_{tx}(\omega,z) A_x^{(i)}(-\omega,z) - \sum_i\frac{f}{2 e^2} A_x^{(i)}(\omega,z)  A_x^{(i)}{}'(-\omega,z) \right.\nn
&& \left.  + \frac{\ell^2}{4\kappa^2 z^2} g_{tx}(\omega, z)  g_{tx}' (-\omega,z) -\frac{\langle \epsilon \rangle}{2}g_{tx}(\omega,z)  g_{tx}(-\omega,z)\right]\,,
\eea
where we have already Fourier transformed the time variable and used the fact that $\Phi(z)\rightarrow 1$ at the boundary. The condition that the solution be asymptotically $AdS$ means that $f(z)\rightarrow 1$ as well. For the configurations we are interested in there is no dependence on the field-theory spatial directions and so the integral $\int d^2x={\cal A}$ will simply give us an overall factor of volume which explains why we included it in the definition \reef{eq:Kubo}.
The fluctuations of the gauge fields and the metric have the near-boundary expansions
\bea
 A_x^{(i)}(\omega,z) &=&  A_x^{(i),\,0}(\omega) + z  A_x^{(i),\,1} + \cdots \,,\nn
 g_{tx} &=&  g_{tx}^{(0)} + \tfrac{z^3}{3}  g_{tx}^{(3)} + \cdots\,.
\eea
Solving these equations with ingoing boundary conditions at the horizon will relate the subleading coefficients to the leading ones, defining the coefficient matrix ${\cal C}$:
\bea\label{eq:connectioncoeffs}
 A^{(i)}_{x,\,1}(\omega) &=&\sum_j {\cal C}_{ij}(\omega)  \, A_x^{(j),\,0}(\omega)  + {\cal C}_{iP}(\omega) \, g_{tx}^0(\omega) \nn
 g_{tx}^{(3)}(\omega)  &=& {\cal C}_{PP}(\omega)   g_{tx}^{(0)}(\omega)  + \sum_j {\cal C}_{P j}\, A_{x}^{(j),\,0}(\omega) \,.
\eea
The matrix ${\cal C}_{ij}$ is symmetric in its indices and time-reversal invariance implies\footnote{Paraphrasing a standard  argument \cite{Herzog:2009xv}: Both $P$ and $J^{(i)}$ are current operators, so they are odd under time reversal. Assuming that the vacuum is invariant under T we must have $G^R_{P J}(\omega) = \eta_J\eta_{P} G^R_{JP}(\omega) =G^R_{JP}(\omega)$.} that ${\cal C}_{Pi} =\frac{2g^2}{\ell^2}C_{iP}$.

The generating functional will give rise to the desired correlation functions of the dual field theory by differentiating with respect to the sources. However, there is still a residual bulk diffeomorphism invariance as the condition $g_{zx}=0$ only partially fixes the gauge. Since we started with a gauge invariant action, we must also have that under the transformation \reef{eq:residualgauge} the function $W$ is invariant:
\be\label{eq:Wvar}
\delta_{\rm res} W[g_{tx}^0,A^{(i),0}_x]=0\,.
\ee
This implies relations between some of the coefficients in \reef{eq:connectioncoeffs}, which have the interpretation of Ward identities. As noted below \reef{eq:residualgauge} we can also think of this transformation as resulting from translations of the dual field theory and so the resulting Ward identities are a consequence of this translation invariance. One finds that \reef{eq:Wvar} implies that
\bea
{\cal C}_{PP} &=& \frac{4\kappa^2}{\ell^2} \langle \epsilon \rangle\,,\nn
{\cal C}_{Pi}  &=& -\frac{\kappa^2}{\ell^2}\langle n^{(i)}\rangle
\eea
Upon differentiating  the on-shell action we obtain
\bea\label{eq.correlators}
\frac{2\pi}{{\cal A}} \langle P(\omega_1) P(\omega_2) \rangle &=& \left(\langle \epsilon\rangle+\langle p \rangle\right)\delta(\omega_1 + \omega_2) \,,\nn
\frac{2\pi}{{\cal A}} \langle P(\omega_1) J^{(i)}(\omega_2) \rangle &=& \langle n^{(i)}\rangle\delta(\omega_1 + \omega_2) \,,\nn
\frac{2\pi}{{\cal A}} \langle J^{(i)}(\omega_1) J^{(j)}(\omega_2) \rangle &=& - \frac{1}{2e^2}\left[ {{\cal C}}_{ij}(\omega_1) +{\cal C}_{ji}(\omega_2)\right]  \delta(\omega_1+\omega_2)\,.
\eea
Differentiating a generating functional fixes the retarded correlation functions up to possible contact terms. We added a contact term proportional to the pressure $\langle p \rangle$ to the first line, which follows from translation invariance \cite{Herzog:2009xv,Hartnoll:2009sz}.  All dynamical information about the conductivities is now encoded in the frequency dependent coefficients ${\cal C}_{ij}(\omega)$. With \cite{Bigazzi:2011ak} let us define the heat current
\be
J_\cQ = P - \sum_i \mu^{(i)}J^{(i)}\,.
\ee
Then the transport coefficients are
\be\label{eq.transportcoeffs}
\sigma_{ij} = \frac{i}{\omega{\cal A}} \langle J^{(i)} J^{(j)}\rangle\,,\qquad T\alpha_{i} = \frac{i}{\omega{\cal A}} \langle J_\cQ J^{(i)}\rangle\,,\qquad T\bar\kappa = \frac{i}{\omega{\cal A}} \langle J_\cQ J_\cQ\rangle\,.
\ee
Here $\sigma_{ij}$ is the conductance matrix, describing both {\it intra}layer and {\it inter}layer conductance. $\bar\kappa$ is the thermal conductivity, which in our system is universal to both layers and finally $\alpha_i$ is the thermoelectric coefficient for each layer. We obtain the following simple  relations between transport coefficients
\bea\label{eq:WardIdentities}
T\bar\kappa &=& \sum_{i,j}\mu^{(i)}\mu^{(j)}\sigma_{ij}  + \frac{i}{\omega} \left(   \langle \epsilon \rangle + \langle p \rangle - 2 \sum_i \mu^{(i)}\langle n^{(i)} \rangle \right) \nn
T\alpha_i &=&\frac{i}{\omega} \langle n^{(i)} \rangle- \mu^{(j)}\sigma_{ji}
\eea
The identities \reef{eq:WardIdentities} generalise the relations of  \cite{Herzog:2007ij} to the double-layer case, albeit without considering magnetic fields. Clearly it would be interesting to extend our results to include them. Care must be taken when comparing to the discussion in \cite{Herzog:2009xv} as the $B\rightarrow 0$ and $\omega\rightarrow 0$ limits do not commute. We shall now analyze the transport coefficients \reef{eq.transportcoeffs}   at low frequencies using the memory matrix approach, which we briefly review.
 
\section{Low-frequency limit and momentum relaxation\label{sec.momentum_relax}}
As mentioned several times already a translation-invariant system at finite charge density has a delta function divergence in its dc electrical conductivity. It is convenient to separate this divergence from the regular part, by writing the conductivity matrix as
\be\label{eq.regularsigma}
\sigma^{ij}(\omega,T) = D^{ij}(T)\left(\frac{i}{\omega} + \pi\delta(\omega) \right) + \sigma_{\rm reg}^{ij}(\omega,T)\,.
\ee
We refer for $D^{ij}(T)$ as the Drude weight, since it becomes  the area under the Drude peak once translation invariance is broken.  This divergence is associated with the conserved  momentum. In the case at hand the divergent part of the conductivity is given by
\be\label{eq.dclimit}
D^{ij}(T) =\frac{\langle n^{(i)}\rangle\langle n^{(j)}\rangle}{\langle \epsilon \rangle+\langle p \rangle} \,,
\ee
so that each matrix element of $\sigma^{ij}$ has a delta function divergence at zero frequency. The more precise reason for this divergence is that each current $J^{(i)}$ has a non-vanishing overlap with the momentum operator, that is $\langle J^{(i)}P \rangle \neq 0$. This implies that a finite-current configuration carries a finite momentum $\langle P  \rangle = \frac{\chi_{J_iP}}{\chi_{PP}}\langle J_i \rangle$. Since momentum is conserved this component of the current `parallel' to $P$ persists at arbitrarily late times and results in the delta-function divergences in \reef{eq.dclimit}. If translation is weakly broken, so that we can think of $P$ as an approximately conserved quantity, the result is a Drude-peak in the conductivity with Drude weight given by $D^{ij}(T)$, as we shall see in more detail below.

Now observe that the Drude weight cancels for precisely the combination $\sigma^{\rm drag}$ introduced in \reef{eq:drag_def}. This is equivalent to the statement that $\langle J_- P\rangle =0$, i.e. that the current mediating $\sigma^{\rm dc}$ has no overlap with the momentum operator. This implies that a current $J_-$ does not have a component `parallel' to the momentum and is thus able to relax on its own intrinsic timescale.  The arguments presented here are somewhat heuristic (although the result \reef{eq.dclimit} is precisely true). We now give a more rigorous derivation of these results using the memory matrix formalism, which will also allow us to make the stronger statement that $\sigma^{\rm dc}$ is independent of the the mechanism of translation breaking, so long as it can be characterized by a coupling of the clean system to the leading irrelevant operator breaking translation. Before we proceed to the calculation itself, it is necessary to introduce a certain amount of formalism, more details of which can be found, for example in \cite{Forster,Hartnoll:2012rj}. We will use the notation of the former reference for the remainder of this paper.
\subsection{Some background on the memory matrix approach}
The memory function approach to transport, to the extent that we use it here, is a reformulation of the usual Kubo approach, suitable for considering late-time effects perturbatively \cite{Mori01031965}. In this approach one develops a perturbative formulation for self-energies in correlation functions, in situations where a perturbation theory of the correlation function itself is ill defined. We introduce, for sake of keeping the paper self contained, the essential notions here.

We start by defining the inner product of two operators $A$ and $B$ at inverse temperature $\beta = 1/T$
\be\label{eq.innerprod}
\Bigl(A\Bigr| B\Bigr) =\beta^{-1}\int_{0}^{\beta} \langle A^\dagger(0) B(i\lambda)\rangle d\lambda\,,
\ee
where $\langle \,\cdot\,\rangle$ denotes an equilibrium average. We also have the time-dependent correlation function
\be\label{eq.correlationfunction}
C_{AB}(t) = \Bigl(A\Bigr| e^{-i Lt} \Bigr| B\Bigr)\,,
\ee
where $L$ is the Liouville operator, $L{\cal O} = i \left[H,{\cal O}  \right]$. This inner product satisfies the same hermiticity and completeness properties as the standard inner product between a bra and a ket. Denoting a general set of operators by ${\cal O}_n$, we note that the inner product defined here allows one to express their susceptibilities in a simple fashion:
\be
\chi_{mn}(T)  = \beta\Bigl({\cal O}_m \Bigl| \Bigr. {\cal O}_n \Bigr)\,.
\ee
The Laplace transform of \reef{eq.correlationfunction}, denoted $\tilde C_{AB}(\omega)$, is related to the retarded response function. More precisely we can use it to write the Kubo formula \reef{eq:Kubo} in the form
\be\label{eq.memory_Kubo}
\sigma^{ij}(\omega) = \beta \tilde C_{J_iJ_j}(\omega)\,.
\ee
Formal manipulations, similar to the ones involved in deriving the fluctuation-dissipation relations, establish that the Laplace transform of \reef{eq.correlationfunction} can be expressed in the form of a dispersion relation
\be\label{eq.memory_response}
\tilde C_{ij}(\omega) = i T\chi_{im} \left[ \omega \chi_{mn}+ iM_{mn}(\omega) \right]^{-1}\chi_{nj} 
\ee
with
\bea\label{eq.memory_matrix}
M_{mn}(\omega) = \beta \Bigl(\partial_t {\cal O}_m \Bigl|{\cal Q}\frac{i}{\omega-{\cal Q}L{\cal Q}} {\cal Q}\Bigr|\partial_t{\cal O}_n \Bigr)\,,
\eea
where we defined the projector
\be\label{eq.projectorQ}
{\cal Q} =1- \sum_{mn} \Bigl| {\cal O}_m\Bigr)\beta\chi_{mn}^{-1}\Bigl( {\cal O}_n \Bigr|\,.
\ee
So far we have done nothing more than to rewrite correlation functions and to repackage the Kubo function in a different form.
If the sum in \reef{eq.projectorQ} runs over all operators in the theory, then the second term in \reef{eq.projectorQ} is a resolution of the identity and nothing is gained. However, as we shall see momentarily, the whole point of the formalism is that it allows us to focus on only a subset of operators, namely those operators which have a non-vanishing overlap with the (approximately) conserved quantities of the system. In the case at hand, these are all those operators that have a non-vanishing overlap with the momentum operator $P$. One often refers to these as the {\it slow modes}, as they relax on scales much larger than the typical microscopic scale of the theory. One then defines the projector ${\cal Q}$ by summing only slow modes, and the formalism becomes a very useful tool to study the late-time physics.
The purpose of writing \reef{eq.memory_response} is to repackage the relevant physical content of $\tilde C_{ij}(\omega)$ in terms of a different function $M_{mn}(\omega)$. This {\it memory} matrix has a good perturbative expansion at late times, where a perturbative approach to $\tilde C_{ij}(\omega)$ is impractical due to divergences of the form \reef{eq.dclimit}.  The susceptibilities $\chi_{mn}$ in \reef{eq.memory_response} incorporate the information about the fast modes relevant to the late-time response. The form of the correlation function \reef{eq.memory_response} is reminiscent of a dispersion relation and in particle physics language $M(\omega)$ would be a matrix of self energies as alluded to above. We will now bring the formalism introduced in this section to bear on the double-layer system which is the subject of our study.

\subsection{Double-layer memory matrix and DC limit}
Henceforth we assume that the effect of translation breaking is mediated by an irrelevant operator which we add to the original Hamiltonian of the system, causing the non-conservation of the momentum via the Heisenberg equation of motion
\be\label{eq.heisenberg}
H = H_0 +g\delta H\qquad \Rightarrow \qquad \dot P = i [H,P] = i g[\delta H,P]\,.
\ee
This can be achieved either via umklapp scattering off a lattice, in which case  $\delta H =  {\cal O}(k_L)$, with lattice scale $k_L$, or it could be more generally via a disorder potential $\delta H = \int dk V(t,k){\cal O}(k)$. In any case we assume that the coupling $g$ is irrelevant, so that the system still flows to the same infra-red fixed point.

The relevant slow modes, i.e. the modes going into a low-frequency expansion of \reef{eq.memory_Kubo} are the set $\left\{J_1,J_2,P  \right\}$. These then form the set of quantities $\left\{  {\cal O}_n\right\}$ entering the memory matrix \reef{eq.memory_matrix} and projector \reef{eq.projectorQ}.
Thus considering the zero-frequency limit of \reef{eq.memory_Kubo} and  \reef{eq.memory_response} one arrives at the dc expression
\be\label{eq.dcDrude}
\lim_{\omega\rightarrow 0}\sigma^{ij}(\omega) = \frac{\chi_{PJ^{(i)}}\chi_{PJ^{(j)}}}{\chi_{PP}}\frac{1}{\Gamma}\,,
\ee
where we should use our previous results $\chi_{P J^i} = \lim_{\omega\rightarrow 0} \langle P J^{(i)}\rangle$, $\chi_{P P} = \lim_{\omega\rightarrow 0} \langle P P\rangle$  for the susceptibilities\footnote{Comparing with \cite{Herzog:2009xv,Hartnoll:2012rj} there is an order of limits issue: we should take $\omega\rightarrow 0$ first and {\it then} $B\rightarrow 0$. I thank Chris Herzog for correspondence on this point.}, and the scattering rate $\Gamma$ is determined by the $PP$ matrix element of the memory function. Using \reef{eq.heisenberg} this boils down to the two-point function of the translation breaking operator, for example \cite{Hartnoll:2012rj}
\be\label{eq.scattering_scale}
\Gamma = \frac{g^2(k_L) k_L^2}{\chi_{PP}}\lim_{\omega\rightarrow 0} \frac{1}{\omega}{\rm Im} G_{{\cal OO}}(\omega,k_L)\Bigr|_{g=0}\,,
\ee
in the case of umklapp scattering. In the case of random disorder, the inverse rate $\tau=1/\Gamma$ is often referred to as the impurity timescale and the analog of \reef{eq.scattering_scale} is derived by considering some averaging over disorder and thus involves an integral over momentum. The operator ${\cal O}$ would be the density operator $J_t^{(i)}$ for electric impurity scattering or scattering off an ionic lattice.

Substituting the expressions \reef{eq.correlators} for the static correlation functions obtained above into this formula leads to a finite dc conductivity with Drude weight equal to the weight of the delta function in \reef{eq.dclimit}. Moreover, if the translation breaking operator is not present, that is if $g \rightarrow 0$,  then $\Gamma\rightarrow 0$ and the divergence in \reef{eq.dclimit} is recovered. 

\subsection{DC limit of $\sigma^{\rm drag}$\label{sec.secdrag}}

Recall now that we noticed that the matrix element $\sigma^{\rm drag}(\omega)=\sigma_{--}(\omega)$ has a finite dc limit even in the translationally invariant case, and moreover that its value is universal for holographic CFTs at finite density. We now consider the effect of translation breaking on this quantity.
One can use the expression \reef{eq.memory_response} for the this case, but since now the relevant current $J_-$ has no overlap with the approximately conserved momentum there is no advantage in using the memory matrix. Its relaxation is entirely determined in terms of microscopic scales contained in $\chi_{mn}(\omega)$. So in  fact it is easier to go back to the original Kubo function \reef{eq.memory_Kubo}. We find the answer
\be\label{eq.CleanLimit}
\lim_{\omega\rightarrow 0}\sigma_{--}(\omega)= \lim_{\omega\rightarrow 0} \frac{1}{\omega}{\rm Im}G^R_{--}(\omega,k=0)\Bigr|_{g=0}\,,
\ee
to leading order with corrections parametrically suppressed by powers of the irrelevant coupling $g$.
 We recovered the same expression for $\sigma^{\rm dc}$ that was obtained previously in the clean limit. This is perhaps not so surprising since we are explicitly assuming that the effect of the translation breaking is irrelevant in the sense of RG, so one would naively say that this must also imply that the effect on conductivities is negligible. But we wish to remind the reader that even such soft translation breaking has a dramatic effect on charge transport in systems where the current operator overlaps with the almost conserved momentum. This is evidenced by the form of the low-frequency conductivity \reef{eq.dcDrude} for a typical element of the conductance matrix. It is because $\sigma^{\rm dc}$ is chosen such that it has no overlap with the conserved momentum, i.e. $\chi_{J_-P}=0$, that its relaxation is given in terms of the microscopic timescales of the CFT as computed in \reef{eq.CleanLimit}, and not in terms of the hydrodynamic timescales associated with conserved quantities of a typical element of the conductance matrix \reef{eq.dcDrude}.

It is important to emphasize that the arguments in this section have nothing to do with the holographic setup, relying only on the fact that there is no net momentum transfer involved in the transport process. In this sense it seems somewhat similar to the observation of Damle and Sachdev \cite{PhysRevB.56.8714} that charge transport near quantum-critical points is finite at the particle-hole symmetric point. With this in mind one could also check that (the sum of) the Azlamasov-Larkin-type diagrams contributing to a perturbative treatment of drag (see Fig.\ref{fig:diagrams}) for the processes contributing to $\sigma^{\rm dc}$ are insensitive do translation breaking in the sense above.

The absence of a Drude peak in $\sigma^{\rm drag}$ is caused by the vanishing cross-susceptibility, $\chi_{PJ_-}=0$. Let us briefly return to the case where only a $\mathbb{Z}_2$ layer-exchange symmetry remains. Clearly $P$ is even under this $\mathbb{Z}_2$, as is $J_+$, which must be suitably defined as a sum of the two currents. However, $J_-$, being a difference of the two currents, is odd, so that $P$ and $J_-$ cannot mix in correlation functions. This then ensures that the cross susceptibility $\chi_{PJ_-}$ vanishes\footnote{This $\mathbb{Z}_2$ is broken unless $\langle n^{(1)} \rangle=\langle n^{(2)} \rangle$, but one can let the $\mathbb{Z}_2$ swap the densities as well, i.e. define a weighted difference of currents as in \reef{eq:drag_def}. Nevertheless the quick argument here should be read with healthy skepticism  unless the densities are equal.}. Hence again we get a vanishing Drude weight and a finite dc limit of $\sigma^{\rm drag}$.

We conclude this section by returning to the original definition of transconductance in section \ref{sec.transconductance}. Recall that this is related to induced charge flow the passive layer, when applying a current in the active layer. The boundary condition on this measurement is that the current flow in the passive layer is zero. It is a simple matter of algebra to use the general relations $J_i = \sigma^{ij}E_j$ together with the zero-current constraint, to show that
\be
E_2 = \left(  \sigma_{12} - \frac{\sigma_{22}\sigma_{11}}{\sigma_{21}}\right)^{-1} J_1 := \rho^{\rm drag} J_1\,,
\ee
where the second equality defines the drag resistivity coefficient $\rho^{\rm drag}$.
The Drude form \reef{eq.dclimit} implies that the quantity in the round brackets is finite and well-defined in the zero-frequency limit, with the Drude weight canceling between the two factors. The zero-current condition together with the Ward identities, as distilled in $D^{ij}(\omega)$ of Eq. \reef{eq.dclimit}, implies that the delta function contributions cancel. This cancellation is a close kin of the lack of Drude weight in the subtracted heat conductivity $\kappa$, defined by measuring heat flow with a zero-current boundary condition. This was recently investigated in the context of a memory-matrix approach in \cite{Mahajan:2013cja}, where it was pointed out that $\kappa$ is similarly a universal quantity for CFTs. 
\section{Discussion}
Let us conclude by evaluating the main points of this paper and their connection to other works. We reiterated the point that charge transport is a subtle physical process \cite{PhysRevLett.85.1092,Roscharxiv,PhysRevB.68.104401,Hartnoll:2012rj,Donos:2012ra,Mahajan:2013cja} and therefore a careful analysis is needed when looking for universal values in the context of holography. Since charge transport usually involves momentum transport, and since momentum relaxation typically happens at hydrodynamical scales, the low-frequency behaviour of conductance (and thus that of resistance) is governed by hydrodynamical modes. We have seen that the quantity $\sigma^{\rm dc}$ defined in \reef{eq:drag_def} involves charge transport without momentum transport, and as such does not have a Drude weight associated with it. Its dc limit is thus determined by the microscopic scales of the strongly-coupled field theory and we have seen that it is indeed universal. It would be interesting to analyse this quantity using a Boltzmann approach or another suitable technique where one can clearly see the interplay of hydrodynamic and microscopic scales involved. Note that spin conductivity (see e.g. \cite{Musso:2013ija,Hashimoto:2013bna}) is a quantity similar to our drag conductivity and one might hope that similar remarks about dc transport apply there. Going away from the dc value, we argued that the optical conductivity $\sigma^{\rm drag}(\omega)$ is independent of frequency for certain strongly-coupled CFTs at finite density. Recently there emerged numerical \cite{Horowitz:2012ky,Horowitz:2012gs,Donos:2012ra} evidence that strongly-coupled theories with a gravity dual can show robust scaling laws in the mid-IR region of the optical conductivity. Similar results have been obtained analytically by considering massive gravity in the bulk \cite{Vegh:2013sk}, albeit with scaling exponents that depend on unfixed parameters. In each of these cases the starting point was a translationally-invariant setup with a non-zero Drude weight, i.e. a conductivity of the form \reef{eq.dclimit}. It would clearly be interesting in this light to investigate the optical properties of our $\sigma^{\rm drag}(\omega)$ as an example of a system with vanishing Drude weight at leading order, firstly in its own right, and also because it might give us some clues about physical processes in {\it bad metals}, which similarly have zero Drude weight \cite{BadMetals}.

The second half of this paper considered the effect of momentum relaxation on the interlayer conductance at the lowest frequencies within the formalism of Mori's memory matrix. We pointed out that a generalization of the argument in \cite{Hartnoll:2012rj} implies that each component of $\sigma^{ij}(\omega)$ has a Drude weight determined entirely in terms of the Ward identities for translation invariance. This furnishes an alternative derivation of the hydrodynamic form \reef{eq.dclimit} at the lowest frequencies. Since the conductance matrix is the outer product of the charge vector $\langle n_i \rangle$ with itself, there is always a linear combination of conductivities for which the Drude weight cancels and we showed that this combination corresponds precisely to the transport coefficient $\sigma^{\rm drag}$. We also showed that scattering off a lattice as well as weak disorder has a negligible effect on this quantity, so that the clean value persists to the case without translation invariance. It is thus tempting to speculate that $\sigma^{\rm dc}$ is a candidate quantity to characterize intrinsic charge transport of a strongly-coupled critical theory. It seems then to be very desirable to investigate in more detail if certain physical systems, such as graphene heterostrucures, can serve as the basis to fashion the requisite strongly-coupled layer systems to compare to the holographic computation. Of course we should not forget that the Coulomb interaction mediating the drag in real graphene (see for example \cite{DragMechanism}) is in fact instantaneous, whereas our setup is fully relativistic. We  nevertheless hope that this work more generally motivates the study of interlayer effects in strongly-coupled theories via holography, with the possibility of comparing to drag measurements. In this context it would be very interesting to study the transconductance in the presence of an applied magnetic field and include the Hall components of $\sigma^{ij}_{ab}$ \cite{MagnetoDrag}.

\subsection*{Acknowledgements}
It is a pleasure to thank Andrew Green, Sean Hartnoll, Chris Herzog, Diego Hofman, Andreas Karch, Elias Kiritsis, Leonid Levitov, Hong Liu, Silviu Pufu, John McGreevy, Subir Sachdev, Senthil, David Tong and Ben Withers for enjoyable and helpful discussions and correspondence. I especially thank Silviu for prompting me to think about the issue of dilatonic fields. This work was initiated at the Simons Symposium `Quantum entanglement: from quantum matter to string theory' at Caneel Bay. I acknowledge the hospitality of the following institutions during the course of this work: The Banff Center, the Texas A$\&$M Mitchell Institute, the Simons Foundation (USVI), and Simons Coffee (Harvard Square). This work was supported in part by the U.S. Department of Energy (DOE) under cooperative research agreement Contract Number DE-FG02-05ER41360.

\appendix
\section{Holographic renormalization \label{app.holorenorm}}
Let us now be more specific about the holographic renormalization of the backgrounds we used implicitly in the main part of this paper. We gave a complete description of the fluctuation part there and ignored the dilaton field $\varphi$ as it does not play a role in the fluctuations. It does however play an indirect role via its backreaction and thus its influence on the thermodynamics and conserved charges. With this appendix we file the required paper work. Let's assume that the potential $V(\varphi)$ has quadratic term $m^2$. Then standard results mean that it is dual to an operator of conformal dimension
\be
m^2 - \Delta_\varphi(\Delta_\varphi-3)=0\,.
\ee
We require that the operator be normalizable, so that $0<\Delta \leq 3$, i.e. that the dual operator be relevant - or at most marginal.
Its near-boundary behavior is then given by the expansion
\be
\varphi(z) = \varphi_1z^\Delta\left(1 + \cdots \right) + \varphi_2 z^{3-\Delta}\left( 1 + \cdots \right)
\ee
with two integration constants $\varphi_1$ and $\varphi_2$. The naive on-shell action then diverges and we must add a local counterterm
\be
S^\varphi_{\rm CT} =  \frac{3-\Delta}{2\kappa^2\ell}\int \sqrt{\gamma} \varphi^2\,.
\ee
If the conformal dimension is in a certain window \cite{Klebanov:1999tb} there is an alternative quantization scheme, which amounts to adding a second counterterm and swapping the definition of source and expectation value, but we shall ignore this issue here.
We can now compute the holographic stress tensor by varying the entire action \reef{eq.counter} with respect to the boundary metric $\gamma_{\mu\nu}$. The result is
\be
\kappa^2\langle T_{\mu\nu}\rangle = \lim_{z\rightarrow 0} \left({\cal K}_{\mu\nu} - {\cal K}\gamma_{\mu\nu} - \frac{2}{\ell}\gamma_{\mu\nu} - \frac{3-\Delta}{4\ell^2} \varphi^2\gamma_{\mu\nu}\right)\,,
\ee
where ${\cal K}_{\mu\nu}$ is the extrinsic curvature of a constant $z$ hypersurface (with outward pointing unit normal) and ${\cal K}$ is its trace.
One finds that
\be
\kappa^2\langle T^\mu{}_\mu\rangle = \frac{3}{2}(\Delta-3) \varphi_1 \varphi_2\,,
\ee
so that the theory is conformal in the UV if the product $\varphi_1 \varphi_2$ vanishes, or if the operator is exactly marginal, i.e. $\Delta=3$. In that case the trace of the stress tensor vanishes and the pressure is half the energy. We find it convenient to define
\be
\lim_{z\rightarrow 0}\frac{\ell}{z} T_{00} = \langle\epsilon \rangle - \frac{3-\Delta}{2\kappa^2}\varphi_1\varphi_2
\ee
with
\be
\langle \epsilon  \rangle = \frac{2 \ell^2}{\kappa^2}\lim_{z\rightarrow 0} \frac{f(1-\sqrt{f})}{z^3}
\ee
and hence the pressure
\be
\langle p \rangle = \frac{\langle \epsilon \rangle}{2} - \frac{3-\Delta}{2\kappa^2}\varphi_1\varphi_2\,.
\ee
Again, we see that in the conformal case the pressure is determined in terms of the energy by the zero-trace condition on the stress tensor.

\bibliographystyle{utphys}
\bibliography{coulombdrag}{}

\end{document}